\documentclass[conference]{IEEEtran}
\IEEEoverridecommandlockouts
\usepackage{cite}
\usepackage{amsmath,amssymb,amsfonts}
\usepackage{algorithmic}
\usepackage{graphicx}
\usepackage{textcomp}
\usepackage{xcolor}
\usepackage{booktabs} 
\usepackage{tabulary}
\usepackage{amsmath,amsfonts,amssymb}
\usepackage{footnote}
\makesavenoteenv{tabular}
\makesavenoteenv{table}
\usepackage{subcaption}
\usepackage{enumitem}
\usepackage{romannum}
\usepackage{authblk}
\usepackage{hyperref}

\newtheorem{mydef}{Definition}
\newcommand*{\affaddr}[1]{#1} 
\newcommand*{\affmark}[1][*]{\textsuperscript{#1}}
\newcommand*{\email}[1]{\texttt{#1}}

\def\BibTeX{{\rm B\kern-.05em{\sc i\kern-.025em b}\kern-.08em
    T\kern-.1667em\lower.7ex\hbox{E}\kern-.125emX}}
\begin{document}

\title{Basket Recommendation with Multi-Intent Translation Graph Neural Network}


\author{Zhiwei~Liu\thanks{*Work is done during the internship at Walmart Labs.}\affmark[1], Xiaohan~Li\affmark[1], Ziwei~Fan\affmark[1], Stephen~Guo\affmark[2], Kannan~Achan\affmark[2], and Philip S.~Yu\affmark[1]\\
\affaddr{\affmark[1]Department of Computer Science, University of Illinois at Chicago, IL, USA} \\ \email{\{zliu213, xli241, zfan20, psyu\}@uic.edu}\\
\affaddr{\affmark[2]Walmart Labs, CA, USA};
\email{\{SGuo, KAchan\}@walmartlabs.com}
}
\IEEEpubid{978-1-7281-6251-5/20/\$31.00~\copyright2020 IEEE}

\maketitle

\begin{abstract}

The problem of basket recommendation~(BR) is to recommend a ranking list of items to the current basket. Existing methods solve this problem by assuming the items within the same basket are correlated by one semantic relation, thus optimizing the item embeddings. However, this assumption breaks when there exist multiple intents within a basket. For example, assuming a basket contains \{\textit{bread, cereal, yogurt, soap, detergent}\} where \{\textit{bread, cereal, yogurt}\} are correlated through the ``breakfast'' intent, while \{\textit{soap, detergent}\} are of ``cleaning'' intent, ignoring multiple relations among the items spoils the ability of the model to learn the embeddings. To resolve this issue, it is required to discover the intents within the basket. However, retrieving a multi-intent pattern is rather challenging, as intents are latent within the basket. Additionally, intents within the basket may also be correlated. Moreover, discovering a multi-intent pattern requires modeling high-order interactions, as the intents across different baskets are also correlated. To this end, we propose a new framework named as \textbf{M}ulti-\textbf{I}ntent \textbf{T}ranslation \textbf{G}raph \textbf{N}eural \textbf{N}etwork~({\textbf{MITGNN}}). MITGNN models $T$ intents as tail entities translated from one corresponding basket embedding via $T$ relation vectors. The relation vectors are learned through multi-head aggregators to handle user and item information. Additionally, MITGNN propagates multiple intents across our defined basket graph to learn the embeddings of users and items by aggregating neighbors. Extensive experiments on two real-world datasets prove the effectiveness of our proposed model on both transductive and inductive BR. The code\footnote{https://github.com/JimLiu96/MITGNN} is available online. 
\end{abstract}


\section{Introduction}\label{sec:intro}
\begin{figure}
    \centering
    \includegraphics[width=0.95\linewidth]{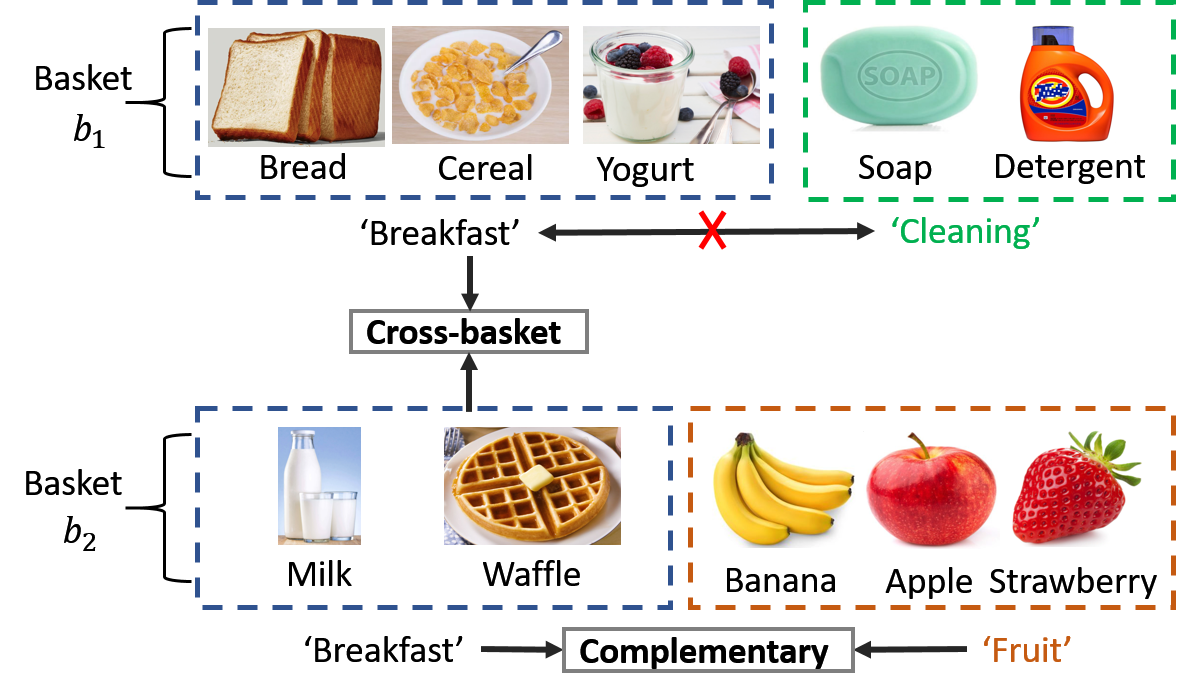}
    \caption{An example of the multi-intent pattern in baskets. ``Breakfast'', ``Cleaning'', and ``Fruit'' are intents in the baskets. Items within the same intents are correlated. Intents may exhibit ``complementary'' relations. The ``Breakfast'' intent relates items across baskets.}
    \label{fig:example}
\end{figure}

The problem of Basket Recommendation~(BR) ~\cite{yu2016dynamic,le2017basket,wan2018representing,liu2020basconv} is to recommend a ranking list of items that users are likely to buy together. It is necessary to model the user-item interactive signal and item-item correlation signal simultaneously~\cite{wan2018representing}. The user-item collaborative filtering~(CF) signal assumes that users with similar interaction patterns share semantically similar items, which is important in representing the user and item semantics~\cite{he2017neural,wang2019neural}. The item-item correlation signal is extracted from items within the same basket~\cite{yu2016dynamic,wan2018representing}, which is crucial for capturing the basket level item relations. For example, \textit{cereal} and \textit{yogurt} are usually in the same basket 
as illustrated in Figure~\ref{fig:example}, as they are complementary with each other~\cite{wan2018representing}. We can model such item relations to recommend a set of items within the same basket. 

Existing methods model item relations in BR by assuming that the items within the same basket are semantically related with each other~\cite{le2017basket,wan2018representing,yu2016dynamic,liu2020basconv,xu2019modeling}. Specifically, Triple2Vec~\cite{wan2018representing} minimizes the distance of the embeddings of items within the same basket. DREAM~\cite{yu2016dynamic} averages the embedding of items within the same basket so as to model the correlations. Xu, et. al. \cite{xu2019modeling} propose to model the items within the same basket as contexts of each other. BasConv~\cite{liu2020basconv} and BGCN~\cite{chang2020bundle} learn basket embeddings by aggregating both user embeddings and item embeddings. These works improve the performance of BR as they can model the basket-level semantics. However, in the real-world, the item relations within baskets are rather complex. According to previous works~\cite{zhou2018deep,li2019graph,zhao2019intentgc,tanjim2020attentive}, users' shopping behavior is usually of multiple interests. Regarding the BR problem, we observe that there are often multiple intents, as illustrated in Figure~\ref{fig:example}. Hence, we should design a new model to handle the multi-intent pattern in the BR problem.


\IEEEpubidadjcol
The advantages of modeling a multi-intent pattern are in three folds. Firstly, the multi-intent pattern helps to find \textbf{complex semantic relations of items}. For example, in Figure~\ref{fig:example}, one basket $b_1$ contains items \{\textit{bread, cereal, yogurt, soap, detergent}\}, where the \{\textit{bread, cereal, yogurt}\} are correlated through the ``breakfast'' intent, while \{\textit{soap, detergent}\} are of ``cleaning'' intent. Simply assuming items within the same basket are of one semantic relation to each other~\cite{wan2018representing,barkan2016item2vec,yu2016dynamic,liu2020basconv} spoils the ability of the model to comprehend the basket-level semantics. Besides, the \textbf{relation between intents} can also be discovered to benefit BR. For example, in Figure~\ref{fig:example}, basket $b_2$ has intents ``breakfast'' and ''fruit'', which are complementary intents~\cite{wan2018representing}, indicating items in one intent should be complements to items in the other. Whereas in $b_1$, the items in ``breakfast'' have no relation to the items in ``cleaning'', though they are in the same basket. As a result, multi-intent relations comprehensively express semantics in baskets. Additionally, the item correlations can be captured more precisely \textbf{across different baskets} through multiple intents. For example, in Figure~\ref{fig:example}, a basket $b_2$ consists of \{\textit{milk, waffle, banana, apple, strawberry}\}, where \{\textit{milk, waffle}\} is of ``breakfast'' intent, while \{\textit{banana, apple, strawberry}\} is of ``fruit'' intent. Though different from $b_1$ in both item-level and basket-level semantics, it still shares a common intent ``breakfast'' with basket $b_1$, and thus we can find the correlation of those items. 

In order to discover the multi-intent pattern from a basket, in this paper, we propose a new framework, \textbf{M}ulti-\textbf{I}ntent \textbf{T}ranslation \textbf{G}raph \textbf{N}eural \textbf{N}etwork~({\textbf{MITGNN}}). Inspired by translation-based models~\cite{bordes2013translating,wang2014knowledge,lin2015learning} in knowledge graph embedding that translate head entity embeddings into tail entity embeddings to model entity relations, MITGNN learns $T$ different translations~\cite{bordes2013translating} from a basket (i.e., head entity) to construct those intents (i.e. tail entities), as illustrated in Figure~\ref{fig:intent_translation}. The translation-based model~\cite{bordes2013translating,wang2014knowledge,lin2015learning} is well-known as it can model the relations between entities. The Translation-based model is built upon triples $(h,r,q)$, where $r$ is the relation between head entity $h$ and tail entity $q$. The embedding of $q$, viewed as the translation of $h$, is learned by adding the embedding of $h$ with the embedding of $r$, i.e., $\mathbf{q} = \mathbf{h} + \mathbf{r}$. Since we model those intents as $T$ tail entities from the same basket entity, we therefore have $\|\mathbf{q}_{i}-\mathbf{q}_{j}\| = \|\mathbf{r}_{i}-\mathbf{r}_{j}\|$. Hence, the usage of translations in MITGNN can also model the relations among multiple intents. For example, in Figure~\ref{fig:intent_translation}, the distance between different intents indicates the correlation of those intents. 

Therefore, MITGNN combines the translation-based model with the GNN model. We construct MITGNN upon our defined basket graph, which consists of user, item, and basket entities. Different from existing methods that learn $T$ relation vectors as free parameters~\cite{bordes2013translating,wang2014knowledge,lin2015learning,he2017translation}, We integrate item and user information into the intent learning process. Inspired by the design intuition of GNN~\cite{hamilton2017inductive,kipf17semi}, the relation vector between basket and intents is learned by $T$-head aggregators. Each head is associated with an intent, aggregating the user and item information of a basket to learn the relation vector. In addition to those intent embeddings, MITGNN learns user embeddings and item embeddings by aggregating from their neighbor embeddings. Since different baskets are connected through user entities and item entities, stacking multiple layers helps to propagate information across different baskets.

The contributions of this paper are:
\begin{itemize} 
    
    \item \textbf{New framework:} We propose a new framework combining the translation-based model with GNN. It can not only propagate information across our defined basket graph, but also model the relations of the learned intents. 

    
    \item \textbf{Multi-intent pattern:} To the best of our knowledge, this is the first work to address the multi-intent pattern in the BR problem. The intents are tail entities translated from basket entities. We use multi-head aggregators to generate the translation vectors. 

    
    \item \textbf{Extensive Experiments:} For the evaluation of our model, we conduct BR experiments on two large-scale real-world datasets. We evaluate the effectiveness of our proposed model under both transductive and inductive BR settings. 
    
    
\end{itemize}

\section{Related Work}\label{sec:related_work}
\subsection{Basket Recommendation~(BR)}
Basket Recommendation~(BR) requires not only user-item CF signals~\cite{he2017neural,zheng2018spectral,wang2019neural}, but also item-item relationships~\cite{mcauley2015inferring,wan2018representing}, e.g., the complementary and substitution relationships. Item A is a substitute for item B if A can be purchased instead of B,
while item A is complementary to item B if it can be purchased in addition
to B~\cite{mcauley2015inferring}. Both of these concepts are extensively investigated in previous work~\cite{mcauley2015inferring,wan2018representing,xu2019modeling}. Sceptre~\cite{mcauley2015inferring} is proposed to model and predict relationships between items
from the text of their reviews and the corresponding descriptions. Item2vec~\cite{barkan2016item2vec} learns item embeddings from user-generated item sets, i.e., the baskets, based on the word2vec model. BFM~\cite{le2017basket} and Prod2vec~\cite{grbovic2015prod2vec} learn one more basket-sensitive embedding for each item, rather than only one embedding, which can help find item relations within the basket. DREAM~\cite{yu2016dynamic} proposes to use RNN structure to exploit the dynamics of user embeddings so as to complete BR. Triple2vec~\cite{wan2018representing} improves within-basket recommendation via (user, item A, item B) triplets sampled from baskets. Later, Xu, et. al. ~\cite{xu2019modeling} propose a Bayesian network to unify the context information and high-order relationships of items. BasConv~\cite{liu2020basconv} is the first work using heterogeneous graph embeddings to complete BR. All these works simply aggregate the information within the basket, while having no discussion towards  basket-level intent mining and exploring little regarding the correlations of those intents.

\subsection{GNNs in recommender systems}
Currently, Graph Neural Networks (GNNs) have been widely explored to process graph-structure data~\cite{liu2020alleviating,dou2020enhancing, liu2020heterogeneous}. Motivated by convolutional neural networks, Bruna et al.~\cite{bruna2013spectral} propose graph convolutions in the spectral domain. Then, GCN~\cite{kipf17semi} simplified the previous graph spectral convolution operations. GraphSAGE~\cite{hamilton2017inductive} inductively generates node embeddings via  sampling and aggregating neighbors. GNNs also have proven their effectiveness in recommender systems. GCMC~\cite{berg2017graph} completes the user-item matrix based on spectral GCN~\cite{kipf17semi}. PinSage \cite{ying2018graph} introduces GraphSAGE \cite{hamilton2017inductive} into a recommender system on an item-item graph. Spectral CF~\cite{zheng2018spectral} leverages spectral convolutions over the user-item bipartite graph. NGCF~\cite{wang2019neural} hierarchically propagates user and item embeddings to model high-order connectivity. DGCF~\cite{li2020dynamic} applies dynamic graph to recommender systems. BasConv~\cite{liu2020basconv} is the first work that uses a GNN to solve the BR problem. Although these methods achieve significant improvements on some types of task, all ignore the multi-intent pattern in the BR problem. Our proposed MITGNN combines a translation-based model with a GNN model to handle the multi-intent pattern.

\subsection{Intent Mining}

Recently, a variety of intent mining methods have been proposed. Temporal Deep Structured Semantic Model (TDSSM) \cite{song2016multi} combines users' long-term and short-term preferences to generate their real-time intention. Deep Interest Network (DIN)~\cite{zhou2018deep} considers the diversity of user interests, and utilizes the attention mechanism to calculate the relevance between the current advertising commodity and historical commodities clicked by the user. Graph Intention Network (GIN)~\cite{li2019graph} applies a graph structure to deal with the behavior sparsity problem.  IntentGC~\cite{zhao2019intentgc} learns explicit preferences and heterogeneous relationships from user behaviors and item information via a GCN. ASLI~\cite{tanjim2020attentive} models users' intents by leveraging their historical interactions, using a self-attention layer and a temporal convolutional network layer to learn the intent embedding. However, regarding the intent mining in BR problems, all existing methods require improvements. As illustrated in Figure~\ref{fig:example}, there are multiple intents in a basket, these intents are composed of several items, and they are correlated with each other. Therefore, we should not only use the combination of items to generate multiple intents, but also model the correlations between intents. 



\section{Preliminary}~\label{sec:preliminary}
    

In this section, we present the preliminaries and definitions related to our Basket Recommendation~(BR) problems. 

\subsection{Problem Definition}
We have a set of items $\mathcal{I}=\{i_{1},i_{2},\dots, i_{|\mathcal{I}|}\}$ and a set of users $\mathcal{U}=\{u_{1}, u_{2}, \dots, u_{|\mathcal{U}|}\}$. The BR problem is defined under two different settings based on whether the basket to predict is given during training, i.e. the transductive and inductive BR problems:
\begin{mydef}
\textbf{(Transductive and Inductive BR):} Assume a set of shopping baskets $\mathcal{B}=\{b_1, b_2, \dots, b_{|\mathcal{B}|}\}$, each of which contains a set of items $\mathcal{I}_{b} \subset \mathcal{I}$ and is associated with a user $u \in \mathcal{U}$. We recommend a ranking list of items ($i^{*} \in \mathcal{I} \setminus \mathcal{I}_{b^{*}}$) to complete the current basket $b^{*}$. If $b^{*}\in \mathcal{B}$, it is a transductive BR problem. Otherwise, if $b^{*}\notin \mathcal{B}$, it is an inductive BR problem.
\end{mydef}



\begin{figure}
    \centering
    \includegraphics[width=0.8\linewidth]{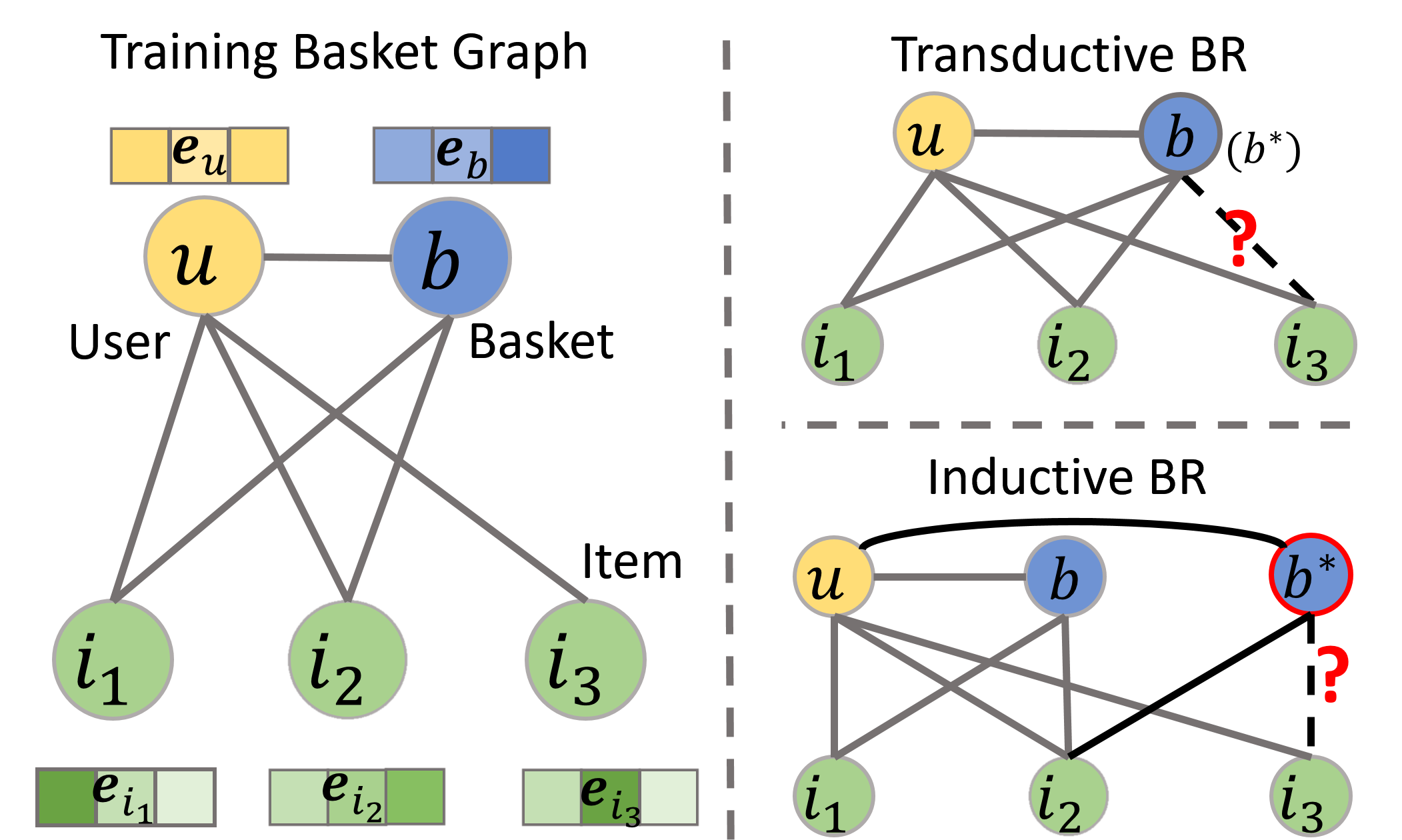}
    \caption{An example of the basket graph for training (left). The prediction process of transductive BR and inductive BR (right).}
    \label{fig:basket_graph}
\end{figure}

The goals of the transductive and inductive BR are both to fill the current basket $b^{*}$ with items not contained in it. However, transductive BR differs from inductive BR in whether the current basket is in the given basket set, or in other words, whether it is in the training data. We present the prediction process of transductive BR and inductive BR on the right-hand side of Figure~\ref{fig:basket_graph}. For transductive BR, $b^{*}=b$ and is already in the training data, while for inductive BR, we have a new basket $b^{*}$. The transductive BR is the same as \textit{within-basket} recommendation~\cite{wan2018representing,liu2020basconv}, and the inductive one is similar to \textit{next-basket} recommendation ~\cite{yu2016dynamic,wan2018representing}. The reason why we use `transductive'(`inductive') rather than `with-in'(`next') will be introduced later. 

During an online shopping session, we find that `add-to-basket' are \textit{asynchronous} behavior of users. Users may add items to the current basket at anytime and save them until checkout. During this period, we can recommend items to the current basket before it is finalized. It suggests that we can use the current basket as the training data, i.e., to make a transductive BR. Additionally, as a result of the latency and infrastructure constraints, it should be better to only consider finalized baskets as training data, and test the BR performance on current baskets, which is inductive BR. Evaluation on both settings contributes to comprehensive analysis of the BR model.

\subsection{Problem Formulation}

We solve the BR problem based on a novel graph structure, i.e. the basket graph, which is as follows:
\begin{mydef}
\textbf{(Basket Graph):} A basket graph $\mathcal{G}$ is defined as $\mathcal{G}= (\mathcal{V}_u, \mathcal{V}_b, \mathcal{V}_i, \mathcal{E}_{ub}, \mathcal{E}_{bi}, \mathcal{E}_{ui} )$, which contains $N$ users, $S$ baskets, and $M$ items. $\mathcal{V}_u$, $\mathcal{V}_b$ and $\mathcal{V}_i$ represent the nodes of user, basket, and item, respectively. $\mathcal{E}_{ub}$, $\mathcal{E}_{bi}$ and $\mathcal{E}_{ui}$ denote the edges between users and baskets, between baskets and items, and between users and items, respectively. 
\end{mydef}

Note that each basket may be connected with more than one item, but only with one user. An example of a basket graph is illustrated on the left-hand side in Figure~\ref{fig:basket_graph}. We can learn the graph embedding based on the defined basket graph as shown on the left-hand side of Figure~\ref{fig:basket_graph}. Then, the basket recommendation problem shifts to predicting the edges between the basket nodes and item nodes, which is presented on the right-hand side of Figure~\ref{fig:basket_graph}. For the current basket $b^{*}$, we rank the scores of predicting the edges $\{f(v_{b^{*}}, v_{i}) | i\in \mathcal{I} \setminus \mathcal{I}_{b^{*}}\}$, where $f(\cdot,\cdot)$ is the learned predictive function of edges. The ranked scores are the returned results of our basket recommender system. We explain the reason why we use `transductive'(`inductive') for the BR problem, rather than `with-in'(`next').

In this paper, we need to train the basket node embedding to make a prediction as shown on the right-hand side of Figure~\ref{fig:basket_graph}. Transductive BR is the same as the \textit{within-basket recommendation}~\cite{wan2018representing,liu2020basconv} setting. However, here we use transductive BR to make it comparable with inductive BR. Inductive BR is similar to \textit{next-basket} recommendation~\cite{wan2018representing,yu2016dynamic}. Under these circumstances, the model 
has no information for the predictive basket during the training procedure. Hence, previous methods use the associated user embedding~\cite{wan2018representing,yu2016dynamic} to complete BR for the next basket. However, we argue that the next basket is \textbf{not empty} during prediction. A user opens a new basket by adding an item to the cart. Then, more items are selected into the current basket. For instance, in the Inductive BR of Figure~\ref{fig:basket_graph}, the current basket $b^{*}$ is already connected with item $i_2$. We should use these connected items to make an inference of the embedding of the current basket before prediction. This inference-before-prediction procedure~\cite{hamilton2017inductive} is an inductive procedure. In this paper, we adopt a similar idea as in~\cite{hamilton2017inductive,xu2020inductive} to train aggregators for the inference of basket embeddings.




\section{Proposed Model}\label{sec:model_MITGNN}
\begin{figure*}
    \centering
    \includegraphics[width=0.95\linewidth]{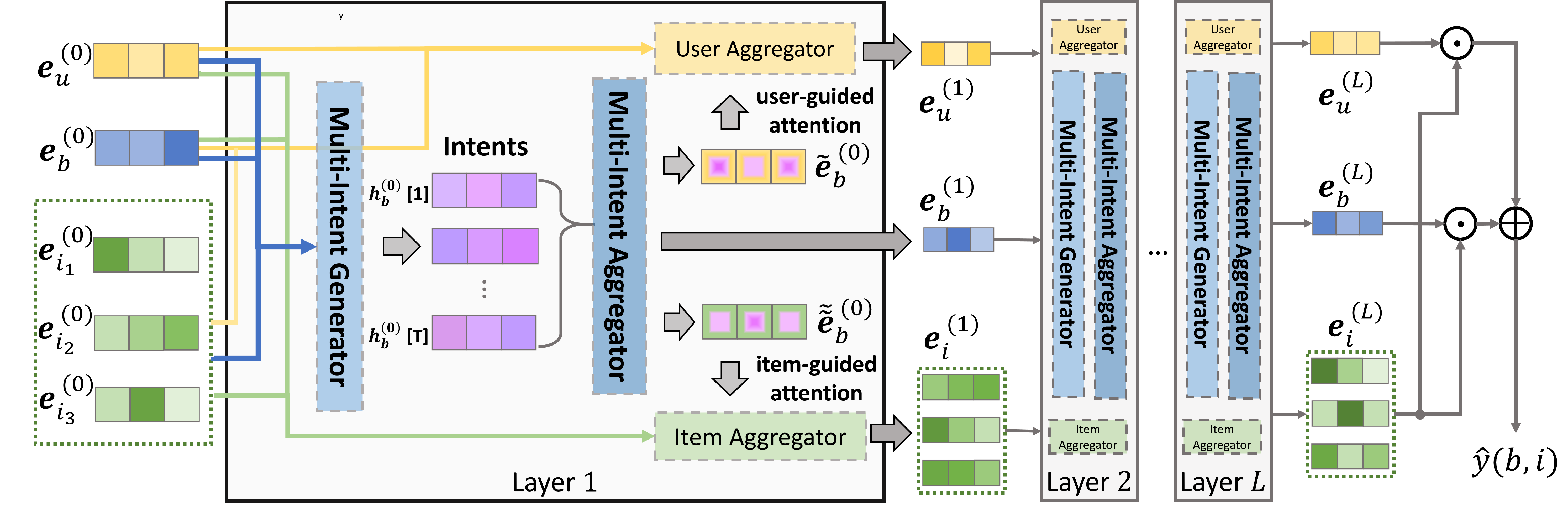}
    \caption{The MITGNN framework. MITGNN is of $L$-layer structure. In the first layer, the multi-intent signals can be captured by the proposed Multi-Intent Generator, which generates $T$ hidden intent $\mathbf{h}_b[t]$ from the input embeddings. Then, the Multi-Intent Aggregator digests intents and outputs three embeddings. One is the basket embedding $\mathbf{e}_b$ for the usage of the next layer. The other two embeddings are the user-guided basket embedding $\Tilde{\mathbf{e}}_b$ and item-guided basket embedding $\Tilde{\Tilde{\mathbf{e}}}_b$ for the usage of the user aggregator and item aggregator, respectively. User and item aggregators output the user and item embeddings, respectively.}
    \label{fig:framework}
\end{figure*}

In this section, we present the structure of our proposed MITGNN model, as illustrated in Figure~\ref{fig:framework}. Our model is based on two designing intuitions: 1) the translation-based model~\cite{bordes2013translating,wang2014knowledge,lin2015learning} that generates multiple intents as tail entities and models correlations of those intents; 2) the GNN framework~\cite{wang2019neural,kipf17semi,hamilton2017inductive} that learns node embeddings by aggregating the information from neighbors.

\subsection{Embedding}
There exist three different types of nodes, i.e., user, item, and basket, which can be represented as $d$ dimensional vector embeddings, denoted as $\mathbf{e}_{u}$, $\mathbf{e}_{i}$, and $\mathbf{e}_{b}$, respectively. Thus, we can define three embedding matrices. For instance, we define the item embedding layer as $\mathbf{E}_i = [\mathbf{e}_{i_{1}}, \mathbf{e}_{i_{2}}, \dots, \mathbf{e}_{i_{|\mathcal{I}|}}]$, where $|\mathcal{I}|$ denotes the total number of items. We retrieve the embedding by the index of the corresponding node. Since the proposed GNN model is an $L$-layer structure, we aggregate the embeddings from one layer to the next layer, and use superscripts to denote the layer number, e.g., $\mathbf{e}_{i}^{(l)}$ represents the embedding of item $i$ at the $l$-th layer. The embeddings of the entities at the $0$-th layer are randomly initialized and trainable. 

\subsection{Multi-Intent Module}
This is one of the main components of MITGNN. The Multi-intent module discovers the multi-intent pattern for baskets. There are two major parts: one is the \textbf{multi-intent generator} that learns multiple intents by modeling them as tail entities translated from basket entities; the other is the \textbf{multi-intent aggregator} that propagates those intents for the usage of next layer. 

\begin{figure}
    \centering
    \includegraphics[width=0.6\linewidth]{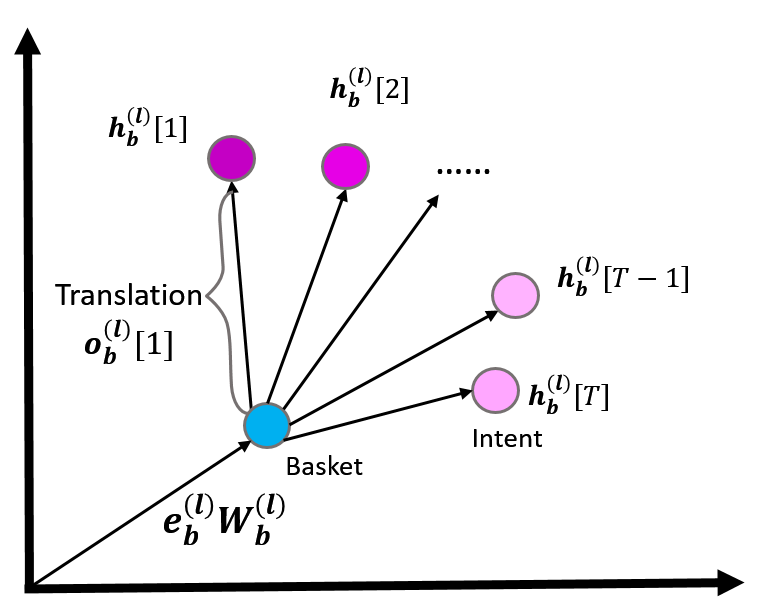}
    \caption{The multiple intents translated from the basket.}
    \label{fig:intent_translation}
\end{figure}

\subsubsection{Multi-intent generator}\label{sec:multi-intent}

Assuming that there are $T$ different intents for the basket $b$, we define a $T$-head intent generator to retrieve $T$ hidden intents from basket $b$. Intuitively, we view the intents as different translations~\cite{bordes2013translating} from the same basket, which is illustrated in Figure~\ref{fig:intent_translation}. The translation-based module is inspired by existing translation-based models~\cite{bordes2013translating,wang2014knowledge,lin2015learning} to learn the embeddings of entities and relations in knowledge graphs. The translation-based module in MITGNN treats $T$ hidden intents as tail entities translated from the head basket entity. Thus, we model the embedding of the $t$-th hidden intent of basket $b$ at $l-$th layer $\mathbf{h}_{b}^{(l)}[t]$, as:
\begin{equation}\label{eq:intent trans}
    \begin{split}
    \mathbf{h}_{b}^{(l)}[t] &= \mathbf{e}_{b}^{(l)}\mathbf{W}_{b}^{(l)} + \mathbf{o}_{b}^{(l)}[t],\quad t=1,2,\dots,T \quad l = 0,1,...,L
\end{split}
\end{equation}
where $\mathbf{e}_{b}^{(l)}\mathbf{W}_{b}^{(l)}$ is mapping the basket information to intent space and $\mathbf{o}_{b}^{(l)}[t]$ denotes the relation vector between the head basket entity and the $t-$th tail intent entity. $\mathbf{W}_{b}^{(l)}\in\mathbb{R}^{d \times d}$ is the weight matrix that propagates the basket embedding from $l-$th layer. The translation $\mathbf{o}_{b}^{(l)}[t]$ models the relation between the basket and $t-$th intent. By using this translation-based model, it is easy to show that the distance (i.e. correlations) of different intents is:
\begin{equation}\label{eq:relation vector}
    \mathbf{h}_{b}^{(l)}[m]-\mathbf{h}_{b}^{(l)}[n] = \mathbf{o}_{b}^{(l)}[m]-\mathbf{o}_{b}^{(l)}[n],
\end{equation}
where $1\leq m,n \leq T$. 

Previous methods~\cite{bordes2013translating,wang2014knowledge,lin2015learning} set the relation vectors $\mathbf{o}_{b}^{(l)}$ as free parameters. However, in BR problem, relation vectors between intents and baskets depend on the information carried by users and items. Thus, we use $T$-head aggregators at different layers to learn it by aggregating neighboring nodes' information. Each $t$-th head is corresponding to $t$-th intent as: 

\begin{equation}\label{eq:basket aggregator}
    \mathbf{o}_{b}^{(l)}[t] =  \mathbf{e}_{u}^{(l)}\mathbf{W}_{1}^{(l)}[t]    \\
    +\sum\limits_{i\in\mathcal{N}_{i}(b)} \mathbf{e}_{i}^{(l)}\mathbf{W}_{2}^{(l)}[t],
\end{equation}
where $\mathbf{e}_{u}$ and $\mathbf{e}_{i}$ denotes the embedding of user and item, respectively. $\mathcal{N}_{i}(b)$ denotes the neighboring items of basket $b$, i.e., the items within the basket $b$. As denoted in Eq.~(\ref{eq:basket aggregator}), each head is associated with a pair of weight matrix $(\mathbf{W}_{1}[t], \mathbf{W}_{2}[t])$, where $\mathbf{W}_{1}[t],\mathbf{W}_{2}[t] \in \mathbb{R}^{d\times d}$. Thus, we only introduce $T$ pairs of weight matrices for generating $T$ intents. $\mathbf{W}_{1}$ is to aggregate the user information, while $\mathbf{W}_{2}$ is to aggregate the item information. 

\subsubsection{Multi-intent Aggregator}\label{sec:multi-intent aggregator}
Since the generated multiple intents are hidden embeddings, those intents should be further aggregated to the next layer to learn node embeddings. Firstly, we should aggregate the multiple intents into the basket embedding for the usage of next layer. As different intents contribute unequally to the basket information, we adopt the idea in attention mechanism~\cite{vaswani2017attention,velivckovic2017graph} and propose an self-attention aggregator for aggregating the embeddings of intents as one basket embedding $\mathbf{e}_b^{(l+1)}$:
\vspace{-2mm}
\begin{equation}\label{eq:basket_aggregator}
{
    \mathbf{e}_b^{(l+1)} = \sigma\left( \sum_{t=1}^{T}\gamma_{bt}^{(l)}\mathbf{h}_{b}^{(l)}[t]\right),
}
\end{equation}
where $\gamma_{bt}^{(l)}$ is the attention weight indicating the importance of $t-$th intent to the current basket $b$ at $l-$th layer, which is:
\begin{equation}\label{eq:intent_att_basket}
{\gamma_{bt}^{(l)} = \frac{\exp\left(\sigma\left(\{\mathbf{h}^{(l)}_b[t]\oplus\mathbf{e}_{b}^{(l)}\}\mathbf{a}_{b}^{\top}\right)\right)}{\sum_{t=1}^{T}\exp\left(\sigma\left(\{\mathbf{h}^{(l)}_b[t]\oplus\mathbf{e}_{b}^{(l)}\}\mathbf{a}_{b}^{\top}\right)\right)}},
\end{equation}
where $\mathbf{a}_{b}\in \mathbb{R}^{2d}$ is the weights for the basket-guide attention layer. The $\sigma(\cdot)$ in Eq.~(\ref{eq:basket_aggregator}) and Eq.~(\ref{eq:intent_att_basket}) denotes the activation function, where we use LeakyReLU\footnote{Other activation function can be used, e.g. Sigmoid.}~\cite{maas2013rectifier}. $\oplus$ denotes the concatenation of two embeddings. Eq.~(\ref{eq:intent_att_basket}) is intuitively the softmax of the information between each intent and the basket. 

Secondly, since GNN framework requires the basket embeddings to learn both user embeddings and item embeddings, the embeddings of those hidden intents of the basket $b$ should be aggregated into one user-guided basket embedding and one item-guided basket embedding. Therefore, we propose to use type-guided attention mechanism to aggregate the intents into two \textbf{type-guided basket embeddings}. The designing intuition is that the correlations between users and intents are distinct from the correlations between items and intents. Because the former is to discover the preference of users to the intents~\cite{liu2017deepstyle, yu2018aesthetic}, while the later is to model the relation of items to the intents~\cite{liu2020basconv}. Hence, we need to apply different weights for each intent vector during intent aggregation. We also use the the self-attention mechanism as type-guided attention like in Eq.~(\ref{eq:basket_aggregator}):
\begin{equation}\label{eq:type-guided embedding}
{
    \Tilde{\mathbf{e}}_b^{(l)} = \sigma\left( \sum_{t=1}^{T}\alpha_{bt}^{(l)}\mathbf{h}^{(l)}_b[t]\right), \quad \Tilde{\Tilde{\mathbf{e}}}_b^{(l)} = \sigma\left(\sum_{t=1}^{T}\beta_{bt}^{(l)}\mathbf{h}^{(l)}_b[t]\right) ,
}
\end{equation}
where $\Tilde{\mathbf{e}}_b^{(l)}$ is the user-guided basket embedding generated by aggregating the hidden intent embedding with user-guided attention weights $\alpha_{bt}^{(l)}$, and $\Tilde{\Tilde{\mathbf{e}}}_b^{(l)}$ is the item-guided basket embedding generated by aggregating the hidden intent embedding with item-guided attention weights $\beta_{bt}^{(l)}$. The weights $\alpha_{bt}^{(l)}$ and $\beta_{bt}^{(l)}$ are calculated from the user-guided and item-guided attention, respectively:
\begin{equation}
{\alpha_{bt}^{(l)} = \frac{\exp\left(\sigma\left(\{\mathbf{h}_b^{(l)}[t]\oplus\mathbf{e}_{u}^{(l)}\}\mathbf{a}_{u}^{\top}\right)\right)}{\sum_{t=1}^{T}\exp\left(\sigma\left(\{\mathbf{h}^{(l)}_b[t]\oplus\mathbf{e}_{u}^{(l)}\}\mathbf{a}_{u}^{\top}\right)\right)}},
\end{equation}

\begin{equation}
{
\quad \beta_{bt}^{(l)} = \frac{\exp\left(\sigma\left(\{\mathbf{h}_b^{(l)}[t]\oplus\Bar{\mathbf{e}}_{i}^{(l)}\}\mathbf{a}_{i}^{\top}\right)\right)}{\sum_{t=1}^{T}\exp\left(\sigma\left(\{\mathbf{h}_b^{(l)}[t]\oplus\Bar{\mathbf{e}}_{i}^{(l)}\}\mathbf{a}_{i}^{\top}\right)\right)},
}
\end{equation}
where $\mathbf{a}_{u}\in\mathbb{R}^{2d}$ and $\mathbf{a}_{i}\in\mathbb{R}^{2d}$ are respectively the weights for user-guided and item-guided attention layer, and $\Bar{\mathbf{e}}_{i}^{(l)}$ is the average of all the item embeddings. Note that the user-guided basket embedding $\Tilde{\mathbf{e}}_b^{(l)}$ and item-guided basket embedding $\Tilde{\Tilde{\mathbf{e}}}_b^{(l)}$ are all intermediate embeddings, which means they are not required to store by the model. By leveraging the user-guided and item-guided basket embeddings, we can further design the user and item aggregators.

\subsection{User and Item Aggregators}
By stacking multiple layers, we propagate the node information to its surrounding neighbors from one layer to the next layer. In other words, the node embedding at $(l+1)-$th layer is aggregated from the $l-$th layer embedding of its neighbors. Previously, we introduce how to aggregate the basket embeddings. In this section, we present user aggregator and item aggregator for learning user and item embeddings. 

\textit{User Aggregator:} Each user is both interacted with several baskets and the corresponding items. Hence, to learn the $(l+1)$-th layer's user embedding $\mathbf{e}_{u}^{(l+1)}$, the user aggregator should sum both the self-information of $u$ and Neighboring Information ($NI_u$) of $u$:

\begin{equation}\label{eq:user_aggregator}
{
    \mathbf{e}_{u}^{(l+1)} = \sigma\left( \mathbf{e}_{u}^{(l)} + 
    NI_{u}^{(l)}
    \right).
}
\end{equation}
Inside the activation function $\sigma(\cdot)$ are two terms. The first term $\mathbf{e}_{u}^{(l)}$ denotes the user information propagated from $l-$th layer. The second term $NI_u^{(l)}$ denotes the Neighbor Information of user $u$ at $l-$th layer, which is:
\begin{equation}\label{eq:user_neighbor_info}
    NI_{u}^{(l)} = \frac{1}{|\mathcal{{N}}_{b}(u)|} \sum\limits_{b\in\mathcal{N}_{b}(u)}\Tilde{\mathbf{e}}_{b}^{(l)}    +\frac{1}{|\mathcal{N}_{i}(u)|}\sum\limits_{i\in\mathcal{N}_{i}(u)} \mathbf{e}_{i}^{(l)},
\end{equation}
where $\mathcal{N}_{b}(u)$ and $\mathcal{N}_{i}(u)$ denote the neighboring baskets and neighboring items of user $u$, respectively. There are two terms on the right hand side in Eq.~(\ref{eq:user_neighbor_info}). The first term is the neighbor information from baskets. The information of basket $b$ is retrieved as multiple intents, which are already aggregated to user-guided basket embedding $\Tilde{\mathbf{e}}_{b}^{(l)}$ as in Eq.~(\ref{eq:type-guided embedding}). Since each user has more than one basket, we average the user-guided basket embeddings for all neighboring baskets.
The second term is the neighboring information from items. We average all the embeddings of the items connected to user $u$.

\textit{Item Aggregator:} Each item is connected by a set of users and also has edges with a set of baskets. Therefore, an item aggregator should sum both the self-information of an item $i$ and the neighboring information $NI_{i}$ of it to learn the item embedding $\mathbf{e}_{i}^{(l+1)}$:
\begin{equation}\label{eq:user_aggregator}
{
    \mathbf{e}_{i}^{(l+1)} = \sigma\left( \mathbf{e}_{i}^{(l)} + 
    NI_{i}^{(l)}
    \right).
}
\end{equation}
Inside the activation function $\sigma(\cdot)$ are two terms. The first term $\mathbf{e}_{i}^{(l)}$ denotes the item information propagated from $l-$th layer. The second term $NI_i^{(l)}$ denotes the Neighbor Information of item $i$ at $l-$th layer, which is:
\begin{equation}\label{eq:item_neighbor_info}
    NI_{i}^{(l)} = \frac{1}{|\mathcal{{N}}_{b}(i)|} \sum\limits_{b\in\mathcal{N}_{b}(i)}\Tilde{\Tilde{\mathbf{e}}}_{b}^{(l)}    +\frac{1}{|\mathcal{N}_{u}(i)|}\sum\limits_{i\in\mathcal{N}_{u}(i)} \mathbf{e}_{u}^{(l)},
\end{equation}
where $\mathcal{N}_{b}(i)$ and $\mathcal{N}_{u}(i)$ denote the neighboring baskets and neighboring users of item $i$, respectively.  There are two terms on the right hand side in Eq.~(\ref{eq:item_neighbor_info}). The first term denotes neighboring information from baskets. The information of basket $b$ is retrieved as multiple intents, which are already aggregated to item-guided basket embedding $\Tilde{\Tilde{\mathbf{e}}}_{b}^{(l)}$ as in Eq.~(\ref{eq:type-guided embedding}). Since each item exists in more than one basket, we average the item-guided basket embeddings for all neighboring baskets. The second term is the neighboring information from users. We average all the embeddings of users connected to item $i$.

By leveraging the user and item aggregators, we can update the embeddings from layer $0$ to layer $L$ and output the final prediction by using the predictive layer.

\subsection{Predictive Layer}
To rank a list of items for recommending a basket, we should output the score between basket $b$ and the items $i$. We obtain the representation of the associated user $u$, the basket $b$, and the item $i$ by inferring and concatenating the embedding at each layer. Thus, the final representation of $u$, $b$, and $i$ respectively are $\mathbf{e}_{u}^{*} = \mathbf{e}_{u}^{(0)} \| \mathbf{e}_{u}^{(1)} \| \dots \|\mathbf{e}_{u}^{(L)}$,  $\mathbf{e}_{b}^{*} = \mathbf{e}_{b}^{(0)} \| \mathbf{e}_{b}^{(1)} \| \dots \|\mathbf{e}_{b}^{(L)}$, and $\mathbf{e}_i^{*} = \mathbf{e}_i^{(0)} \| \mathbf{e}_i^{(1)} \| \dots \|\mathbf{e}_i^{(L)}$. Concatenation of embeddings from all layers stores the information from the initial embeddings to high-order aggregated embeddings. 
They are passed into a predictive function. In this paper, the predictive function is defined as the sum of two dot products:
\begin{equation}
    \hat{{y}}(b,i;u) = \mathbf{e}_{u}^{*}\mathbf{e}_{i}^{*\top} + \mathbf{e}_{b}^{*}\mathbf{e}_{i}^{*\top},
\end{equation}
where the first term $\mathbf{e}_{u}^{*}\mathbf{e}_{i}^{*\top}$ predict the score between user and item, and the second term $\mathbf{e}_{b}^{*}\mathbf{e}_{i}^{*\top}$ predict the score between basket and item. The combination of those two scores measures how likely we should recommend the item $i$ to basket $b$. The user $u$ is the associated user of basket $b$. 

\subsection{Optimization}
We use BPR loss~\cite{rendle2009bpr} to optimize the parameters of our MITGNN model. Given a basket $b$, we sample the positive sample $i$ as the items within the current basket and negative sample $j$ from the other items. Additionally, we add a regularization term for the parameters. Hence, the final optimization loss function is:
\begin{equation}
\mathcal{L} = -\sum_{(b,i,j)\in \mathcal{S}}\log\sigma\big(\hat{{y}}(b,i) - \hat{{y}}(b,j)\big) + \lambda\|\Theta\|_{2}^{2},
\end{equation}
where $\mathcal{S}$ is the set of the generated samples, $\Theta$ denotes all the parameters requiring for training, and $\lambda$ is the regularization factor. Between each layer, we add a normalization layer and dropout-out layer to the embeddings, which stabilizes the training. We optimize the model with the mini-batch Adam~\cite{kingma2014adam} optimizer in Tensorflow. 
\subsection{Inductive Basket Recommendation}~\label{sec:inductive BR}
As we mentioned before, inductive BR is also crucial in real-world applications. Hence, we should evaluate the MITGNN model under the inductive setting. However, under the inductive setting, we do not have basket information until the testing process. In other words, we rank a list of items to a new basket $b^{*}$, but without knowing its existence or which items are in it during training. The problem is that we have no trained embedding of $b^{*}$. Hence, we should make an inference of the embedding of this new basket $b^{*}$. Recalling that in Eq.~(\ref{eq:basket_aggregator}), we learn the basket embedding at the $(l+1)$-th layer by attentively aggregating the embeddings of multiple intents. However, the embeddings of intents and the attention weights require the embedding of the basket at the $l$-th layer, as illustrated in Eq.~(\ref{eq:intent trans}) and Eq.~(\ref{eq:intent_att_basket}). Therefore, we should firstly compute the embedding of $b^{*}$ at the $0$-th layer as:
\begin{equation}
    \mathbf{e}_{b^{*}}^{(0)} = \mathbf{e}_{u}^{(0)} + \frac{1}{|\mathcal{N}_{i}(b^{*})|}\sum\limits_{i\in\mathcal{N}_{i}(b^{*})} \mathbf{e}_{i}^{(0)},
\end{equation}
where $\mathbf{e}_{u}^{(0)}$ is the embedding of the corresponding user at the $0$-th layer, and $\mathbf{e}_{i}^{(0)}$ is the embedding of items within $b^{*}$. We use $\mathcal{N}_{i}(b^{*})$ denotes the items within the new basket $b^{(*)}$. The other variables are already trained and set as constants during testing. Thus, we can calculate the $L$-layer embedding of $b^{*}$ by inferring with Eq.~(\ref{eq:basket_aggregator}), Eq.~(\ref{eq:intent trans}), and
Eq.~(\ref{eq:intent_att_basket}) simultaneously.

\section{Experiments}\label{sec:experiments}

\begin{table*} 
\centering
\vspace{-1em}
\caption{Inductive Basket Recommendation on Instacart}
\vspace{-1em}
\label{tab:next-basket-insta}
\resizebox{0.95\hsize}{!}{
\begin{tabular}{lccccccccc}
\toprule
\textbf{Method} & \text{\small \textit{Recall}@\textbf{20}}   &\text{\small \textit{Recall}@\textbf{60}}  & \text{\small \textit{Recall}@\textbf{100}}  &\text{\small \textit{HR}@\textbf{10}}  & \text{\small \textit{HR}@\textbf{20}}  &\text{\small \textit{HR}@\textbf{30}}  & \text{\small \textit{NDCG}@\textbf{20}}   &\text{\small \textit{NDCG}@\textbf{60}}   & \text{\small \textit{NDCG}@\textbf{100}} \\
\midrule

BPR-MF      & 0.12185 & \underline{0.27980}  & \underline{0.38629} & 0.88981 & \underline{0.97004} & 0.98898 & \underline{0.29311} & \underline{0.46678} & 0.56721 \\
Triple2vec & \underline{0.12532} & 0.27423 & 0.37476 & 0.88326 & 0.96694 & 0.98795 & 0.27538 & 0.45588 & 0.56051  \\
DREAM      & 0.10841  & 0.22782  & 0.30663  & 0.83231  & 0.92101    & 0.95107  & 0.19380  & 0.21817   & 0.25982  \\
GC-MC       & 0.09804 & 0.20555 & 0.27952 & 0.83333 & 0.93974 & 0.97624 & 0.27794 & 0.43145 & 0.51974 \\
R-GCN      & 0.10147 & 0.21788 & 0.29679 & 0.84676 & 0.94180  & 0.97314 & 0.27375 & 0.43339 & 0.52435 \\
NGCF       & 0.12389 & 0.27346 & 0.37504 & \underline{0.89360}  & 0.96832 & \underline{0.99036} & 0.28978 & 0.46565 & \underline{0.56724} \\
MITGNN & \textbf{0.13321} & \textbf{0.30079} & \textbf{0.41234} & \textbf{0.90461} & \textbf{0.97417} & \textbf{0.99277} &  \textbf{0.32173} & \textbf{0.49197} & \textbf{0.59701}  \\
    
\bottomrule
\end{tabular}
}
\end{table*}

\begin{table*} 
\centering
\vspace{-1em}
\caption{Inductive Basket Recommendation on Walmart}
\vspace{-1em}
\label{tab:next-basket-Walmart}
\resizebox{0.95\hsize}{!}{
\begin{tabular}{lccccccccc}
\toprule
\textbf{Method} & \text{\small \textit{Recall}@\textbf{20}}   &\text{\small \textit{Recall}@\textbf{60}}  & \text{\small \textit{Recall}@\textbf{100}}  &\text{\small \textit{HR}@\textbf{10}}  & \text{\small \textit{HR}@\textbf{20}}  &\text{\small \textit{HR}@\textbf{30}}  & \text{\small \textit{NDCG}@\textbf{20}}   &\text{\small \textit{NDCG}@\textbf{60}}   & \text{\small \textit{NDCG}@\textbf{100}} \\
\midrule

BPR-MF     & \underline{0.05339} & 0.12636 & 0.17972 & \underline{0.57442} & 0.76543 & 0.85107 & \underline{0.18325} & \underline{0.32260}  & \underline{0.40903} \\
Triple2vec & 0.04077 & 0.11186 & 0.15998 & 0.39972 & 0.68647 & 0.81179 & 0.12562 & 0.27088 & 0.35430  \\
DREAM      & 0.02913   & 0.06931  & 0.10127  & 0.40463  & 0.59959  & 0.70738  & 0.05675  & 0.06522  & 0.08233  \\
GC-MC      & 0.03041 & 0.08527 & 0.12605 & 0.38272 & 0.61263 & 0.73072 & 0.12383 & 0.25075 & 0.33155 \\
R-GCN      & 0.04133 & 0.11531 & 0.17078 & 0.44631 & 0.70137 & 0.82320  & 0.12500   & 0.26917 & 0.36091 \\
NGCF       & 0.05233 & \underline{0.12651} & \underline{0.18135} & 0.55416 & \underline{0.76660}  & \underline{0.86117} & 0.17709 & 0.31654 & 0.40540  \\
MITGNN  & \textbf{0.05757} & \textbf{0.15062} & \textbf{0.22466} & \textbf{0.61472} & \textbf{0.80992} & \textbf{0.90030}  & \textbf{0.19086} & \textbf{0.33013} & \textbf{0.42062}
\\
    
\bottomrule
\end{tabular}
}
\end{table*}

\subsection{Dataset}

We conduct experiments on two real-world datasets, the \textit{Instacart} dataset and the dataset collected from the \textit{Walmart} online grocery shopping website\footnote{\url{https://www.walmart.com/grocery}}.

\begin{itemize}
    \item \textbf{Instacart} is an online grocery shopping dataset, which is published by \textit{instacart.com}~\footnote{\url{https://www.instacart.com/datasets/grocery-shopping-2017}}. It contains over 3 million grocery transaction records from over 200 thousand users on around 50 thousand items.  
    \item \textbf{Walmart Grocery} is an online service provided by \textit{walmart.com} for shopping groceries. We sampled  $100$ thousand users, whose transaction data are retrieved to conduct the experiment. 
\end{itemize}
After filtering baskets with less than $30$ items and users with less than $5$ baskets for the Instacart data, the final Instacart data contains $65,859$ nodes and $1,271,195$ interactions. We filter baskets with less than $40$ items and users with less than $5$ baskets for the Walmart data, leading to $85,315$ nodes and $1,225,155$ interactions in total. The details of the data statistics are summarized in Table~\ref{tab:data stats}.

\begin{table}
\centering
\vspace{-1em}
\caption{Dataset Statistics}
\vspace{-1em}
\label{tab:data stats}
\begin{tabular}{cccccc}
\toprule
 \textbf{Dataset} & \textit{\small \#Edges} &  \textit{\small \#Baskets} & \textit{\small \#Users} & \textit{\small \#Items}  &  \textit{\small Density} \\ 
\midrule
  \textbf{Instacart}& $1,271,195$  & $32,201$ & $2,904$ & $30,754$ &  $ 0.0293\%$ \\
  \textbf{Walmart} &  $1,225,155$ & $27,797$ & $7,110$ & $50,408$ & $0.0168\%$ \\
\bottomrule
\end{tabular}
\vspace{-2mm}
\end{table}

\subsection{Experimental Setting}
We evaluate performance based on Top-\textbf{K} ranking results, with the evaluation metrics being \textit{Recall}@\textbf{K}, \textit{Hit Ratio}@\textbf{K},  and \textit{NDCG}@\textbf{K}~\cite{resnick1997recommender,he2017neural}. Experiments use \textbf{K}=$\{20,40,60,80,100\}$ for comparison. We first compute the scores between the basket $b^{*}$ and all potential items. Then, we sort the items based on the scores. Finally, we compute the evaluation metrics by comparing the ranking results with the true items within basket $b^{*}$.
The overall comparison is conducted under two settings: one is transductive BR; the other one is inductive BR, as we introduced in Sec.~\ref{sec:preliminary}. We compare MITGNN with the following baselines:
\begin{itemize}[leftmargin=*]
    \item
    \textbf{BPR-MF}~\cite{rendle2009bpr}: we merge the items within the baskets for each user and sample positive and negative items for the user. BPR-MF completes the BR based solely on user embeddings.
    
    \item
    \textbf{Triple2vec}~\cite{wan2018representing}: we sample triplets from baskets and train the user and item embeddings. 
    
    \item
    \textbf{DREAM}~\cite{yu2016dynamic}: it train user embeddings by modeling the dynamics of baskets.
    
    \item
    \textbf{GC-MC}~\cite{berg2017graph}: we adopt the idea in~\cite{berg2017graph} using the GCN model~\cite{kipf17semi} to predict the basket-item relations in the graph. 
    
    \item 
    \textbf{R-GCN}~\cite{schlichtkrull2018modeling}: A GNN model to complete the relations between nodes, which is used here to predict the basket-item relations. 
     
    \item
    
    \textbf{NGCF}~\cite{wang2019neural}: we modify it to predict the basket-item interactions. However, it has no module for the multi-intent pattern.
    
\end{itemize}

For BPR-MF, since it has no basket embedding to evaluate transductive and inductive basket recommendation, we still use the user embedding as the embedding for the next basket. Triple2vec adopts the idea that the basket embedding is the sum of the user embedding and the average of embeddings of items within the basket. 

\begin{figure*}
\begin{subfigure}{.33\textwidth}
    \centering
    \includegraphics[width=\textwidth]{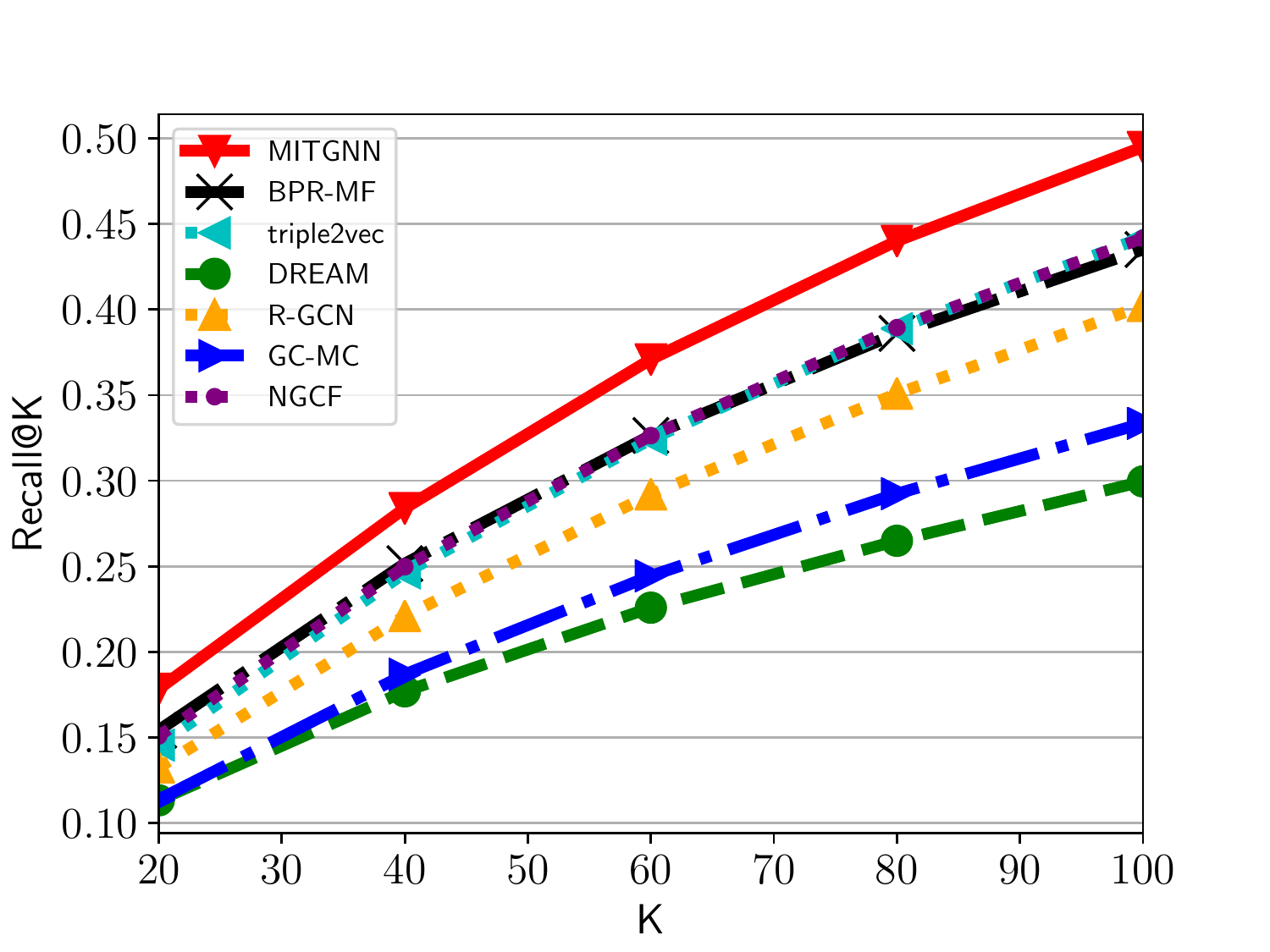}
    \caption{\text{Recall on Instacart}}
    \label{fig:recall_inscart}
\end{subfigure}\hfill
\begin{subfigure}{.33\textwidth}
    \centering
    \includegraphics[width=\textwidth]{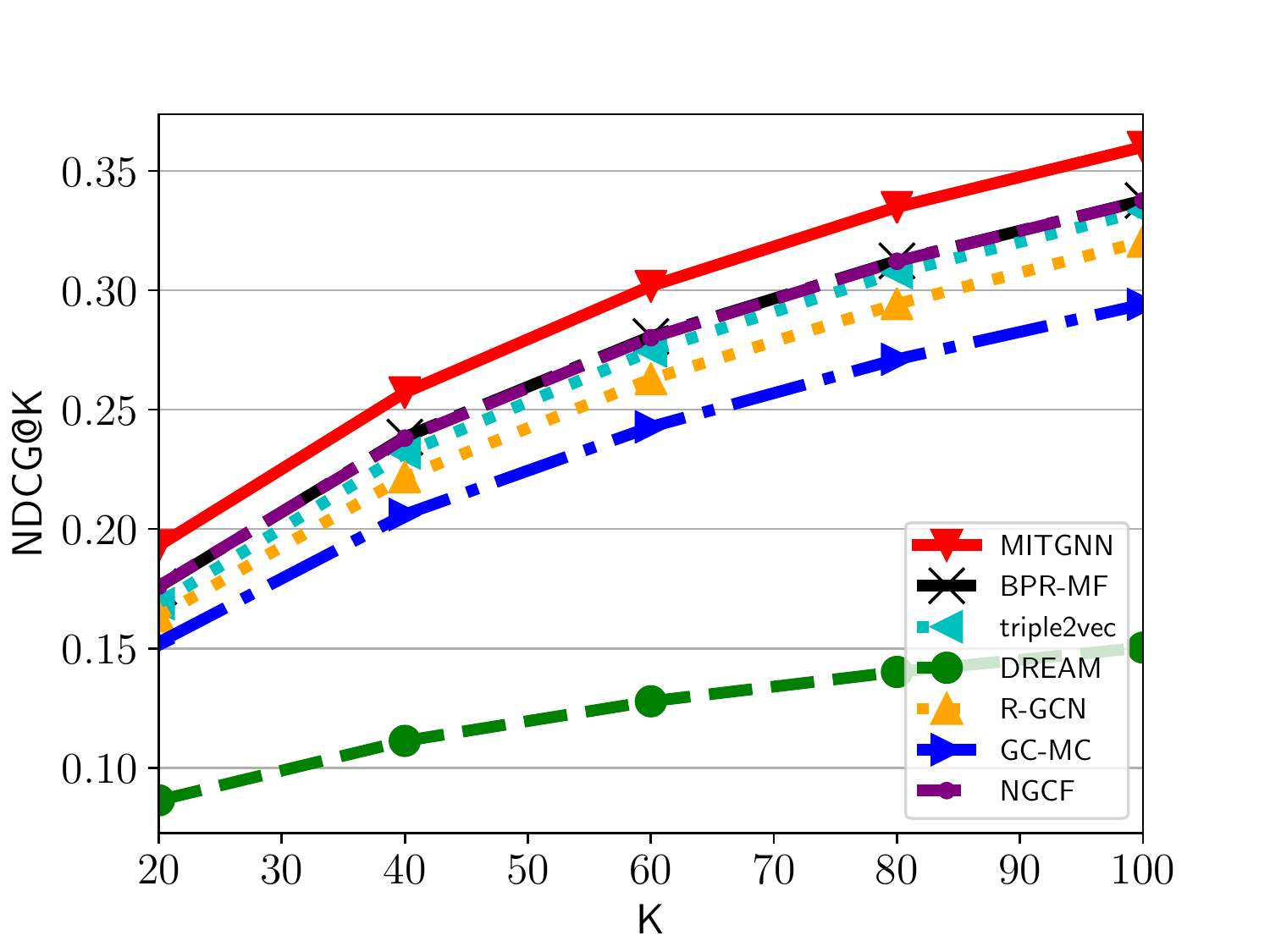}
    \caption{\text{NDCG on Instacart}}
    \label{fig:NDCG_inscart}
\end{subfigure}\hfill
\begin{subfigure}{.33\textwidth}
    \centering
    \includegraphics[width=\textwidth]{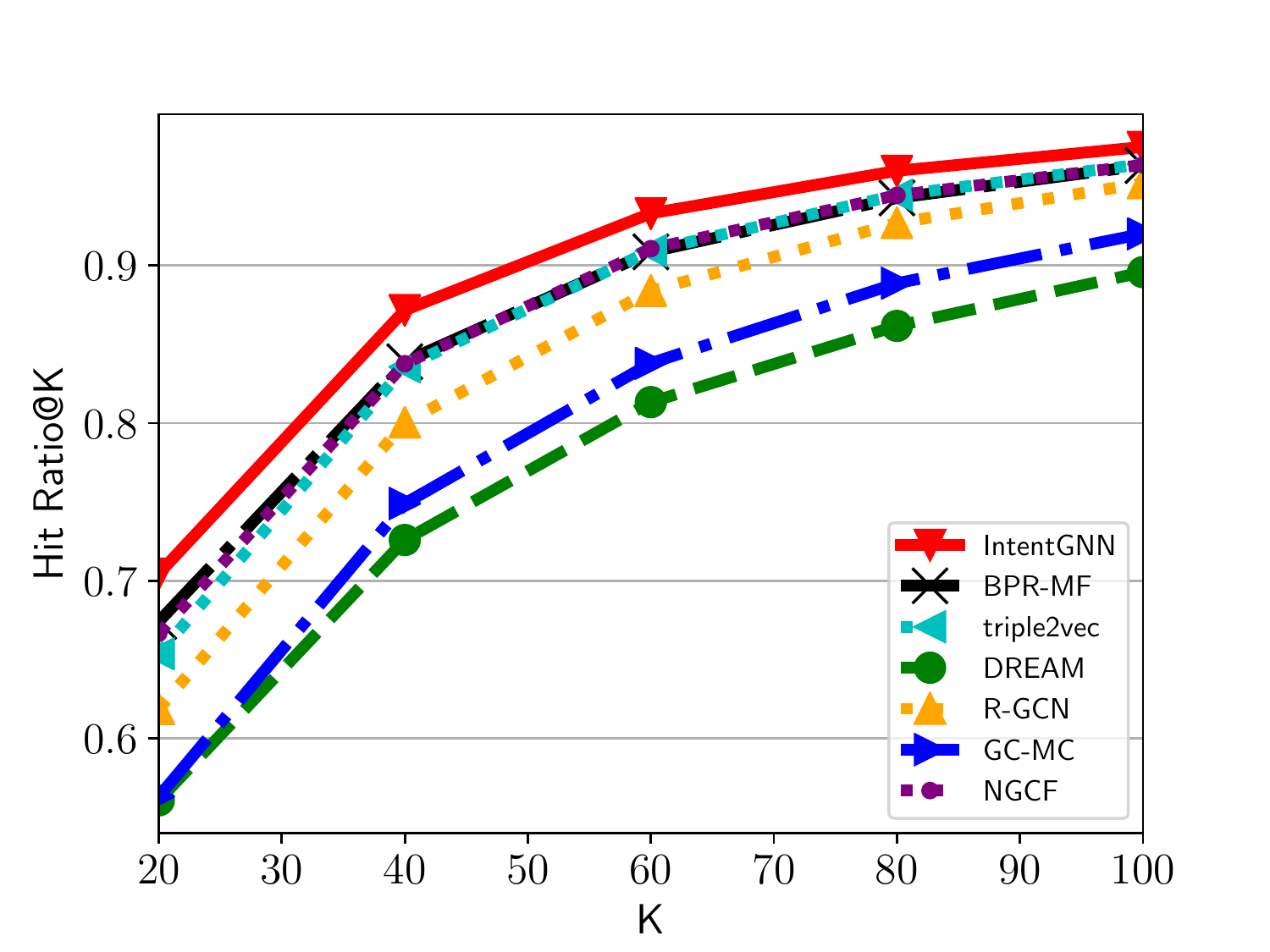}
    \caption{\text{Hit Ratio on Instacart}}
    \label{fig:HR_inscart}
\end{subfigure}
\\
\vspace{-2pt}
\begin{subfigure}{.33\textwidth}
    \centering
    \includegraphics[width=\textwidth]{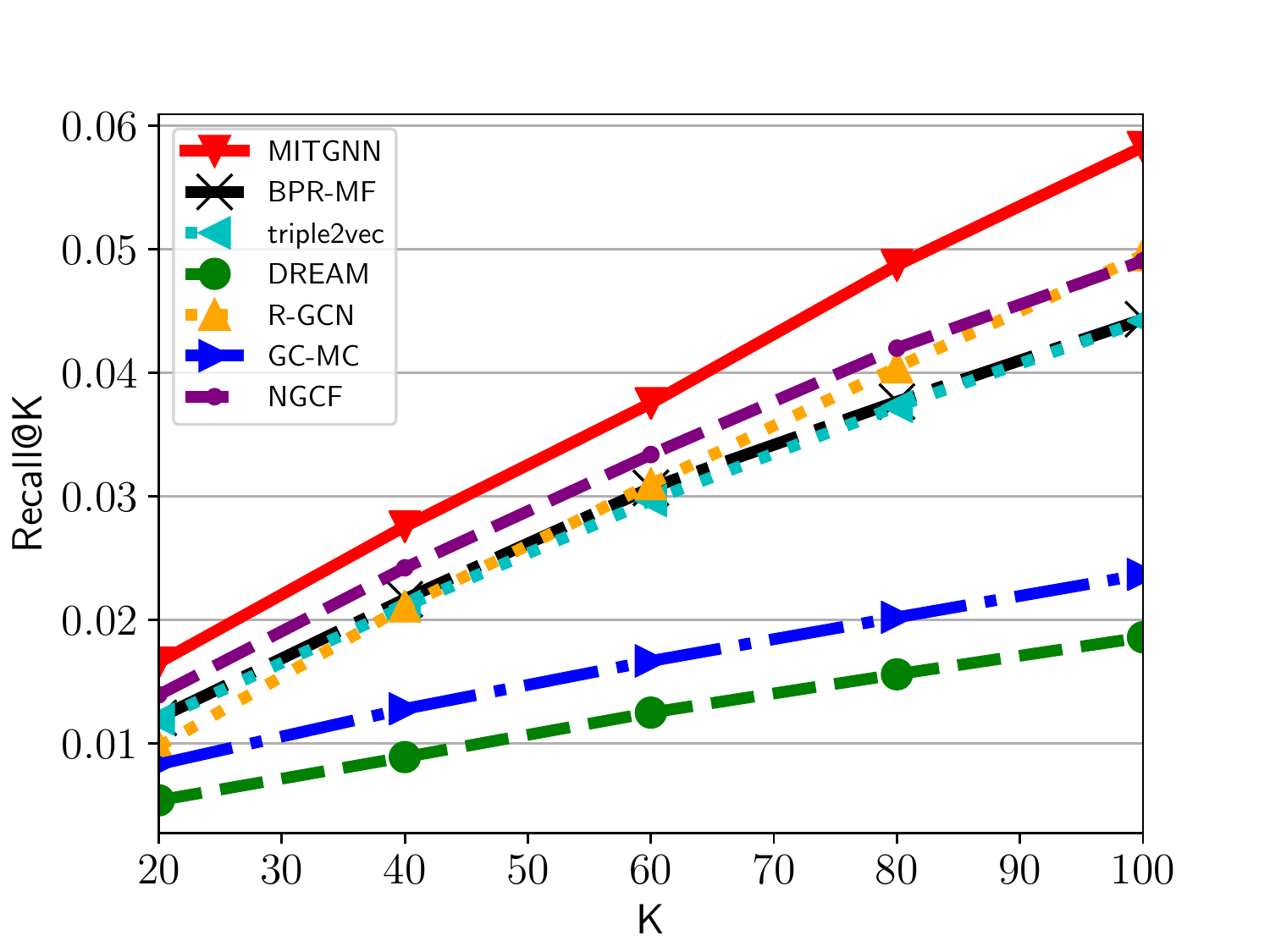}
    \caption{\text{Recall on Walmart}}
    \label{fig:recall_walmart}
\end{subfigure}\hfill
\begin{subfigure}{.33\textwidth}
    \centering
    \includegraphics[width=\textwidth]{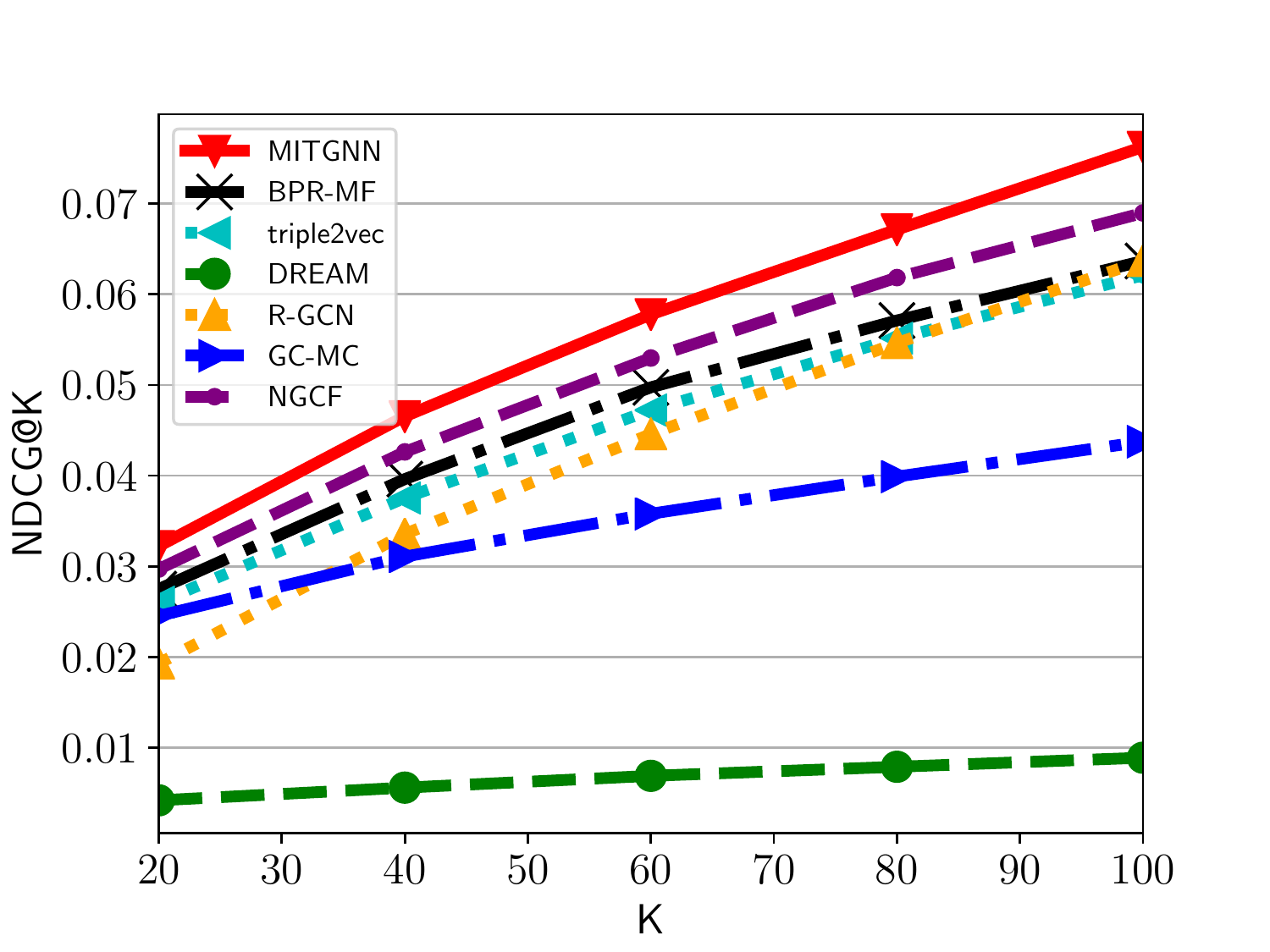}
    \caption{\text{NDCG on Walmart}}
    \label{fig:NDCG_walmart}
\end{subfigure}\hfill
\begin{subfigure}{.33\textwidth}
    \centering
    \includegraphics[width=\textwidth]{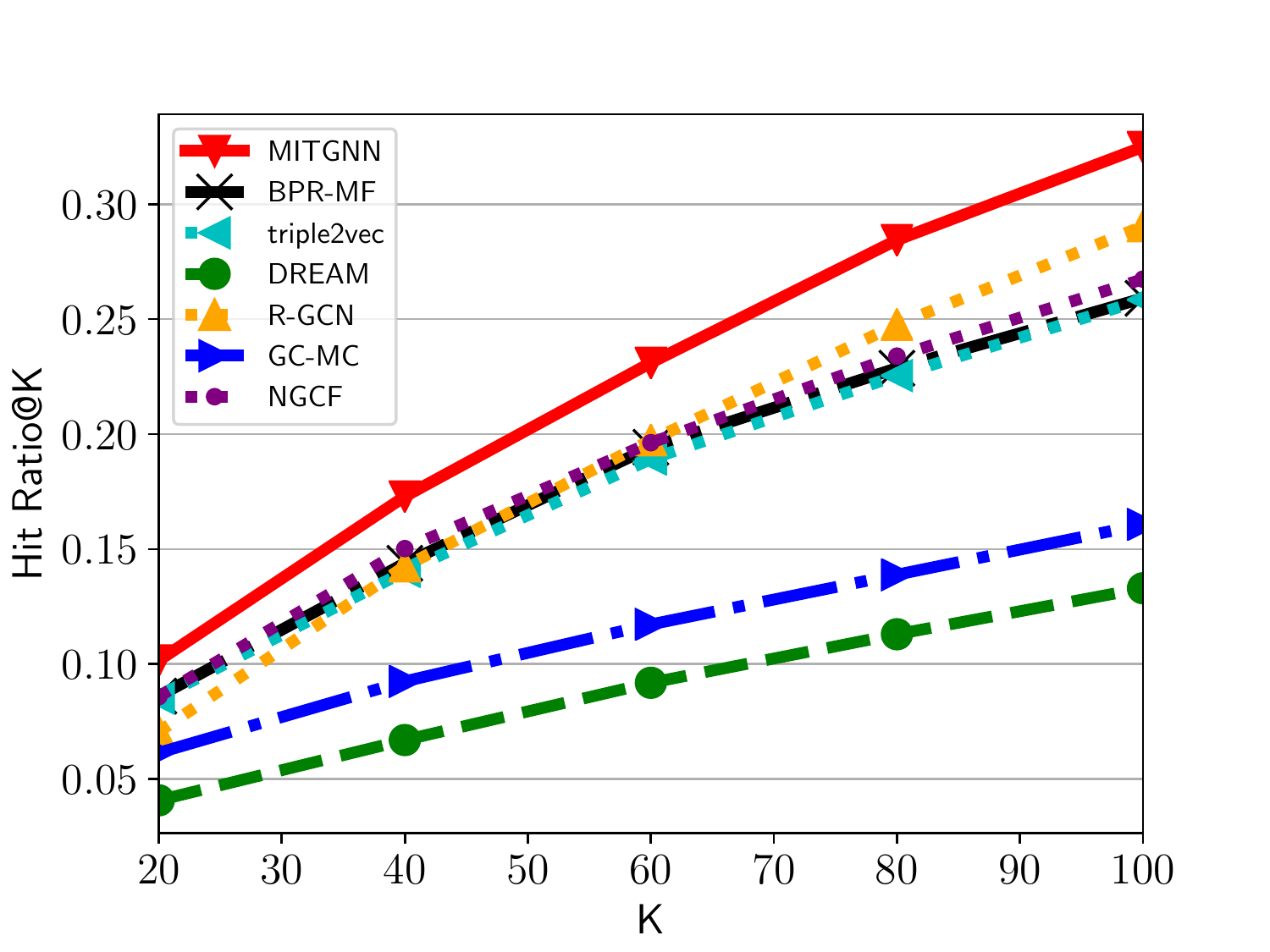}
    \caption{\text{Hit Ratio on Walmart}}
    \label{fig:HR_walmart}
\end{subfigure}\hfill
\caption{Performance of transductive basket recommendation  w.r.t. different \textbf{K} on two datasets.}
\label{fig:transductive}
\end{figure*}

\subsection{Parameter Settings}
All models are validated on the performance of Recall@100. The embedding size $d$ is $64$ for all models. The learning rate for all models is searched from $\{10^{-5}, 5\times10^{-5}, 10^{-4}, 5\times10^{-4}, 10^{-3}, 5\times10^{-3}\}$. The regularization factor $\lambda$ is searched from $\{10^{-5}, 10^{-4}, \dots, 10^{-1}\}$. For the layer number of GCMC, R-GCN, NGCF, and MITGNN, we search the layer number from $\{1,2,3,4\}$. The number of intents for MITGNN is selected from $\{1,2,3,4,5\}$. We choose the MITGNN model with $3$-layer structure for both datasets with learning rate of $0.0005$ for Instacart data and $0.001$ for Walmart data, regularization factor of $0.0001$ for Instacart and $0.00001$ for Walmart data, The number of intents $T$ is tuned as $3$ and $4$ for Instacart and Walmart data, respectively. The details of the tuning process for the hyperparameters are discussed in Sec.~\ref{sec:effect_intent}.

\subsection{Transductive Recommendation}

In this section, we conduct experiments regarding the transductive BR problem. For both datasets, $80\%$ of items within the basket are used as training data. The remaining $20\%$ of items within the basket are used for testing. The performance comparison on two datasets is presented in Figure~\ref{fig:transductive}.   We have the following observations:
\begin{itemize}
    \item MITGNN performs the best against all other baselines. On Instacart data, compared with the strongest baseline on each metric, MITGNN on average improves the Recall, HR, and NDCG by $12.03\%$, $9.43\%$, and $9.71\%$, respectively. On Walmart data, MITGNN on average improves the Recall, HR, and NDCG by $17.82\%$, $17.17\%$, and $10.75\%$, respectively. The improvements suggest that the multi-intent pattern is important in the BR problem.

    \item NGCF outperforms almost any other baseline on both datasets. It aggregates the neighboring information and learns multi-layer embeddings, rather than only using the last layers' embedding like GC-MC and R-GCN. However, R-GCN beats NGCF for \textit{Hit Ratio@}80 and \textit{Hit Ratio@}100, which suggests different relations should also be modeled. MITGNN performs the best as it can model the multi-intent pattern and learn the relations between intents.  
    
    \item Triple2vec and NGCF perform similarly to BPR-MF, which only models the interactions of user-item. This suggests that direct retrieval of item relations from the basket cannot improve the performance by only modeling user-item relations. However, they still outperform other direct aggregation methods such as R-GCN on the Instacart dataset, which indicates that the CF signal and the user factor are important for basket recommendation on the Instacart dataset. On Walmart data, R-GCN performs comparably with BPR-MF and triple2vec, which implies the difference of relations should be modelled. 
    
    \item DREAM performs the worst among all the models as we find that personalized information is more important than sequential patterns in both datasets, which limits the effectiveness of the RNN structure in the DREAM model. 
    
    \item The performance on Instacart is better than that of the Walmart. The first reason is that the Walmart data is much sparser than the Instacart. The former's density is $0.0168\%$, while the latter's density is $0.0293\%$. The second reason is that each user in the Instacart data on average owns $11.1$ baskets. However, compared with only $3.9$ baskets in the Walmart data. The scarcity of personalized information leads to the poor performance.

\end{itemize}

\subsection{Inductive Recommendation}
In this section, we conduct inductive BR experiments. We predict items in the most recent baskets of users, where those baskets are not presented during training. Note that, for the current basket $b^{*}$, we perform inference of the basket embedding $e_{b^{*}}$ as introduced in Sec.~\ref{sec:inductive BR}. We take out $5$ items for aggregation and set the rest as ground truth. During testing, MITGNN learns basket embeddings before prediction. Performance is presented in Table~\ref{tab:next-basket-insta} for Instacart data and Table~\ref{tab:next-basket-Walmart} for Walmart data. Due to the space limitations, we only present the performance of Recall and NDCG@$\{20,60,100\}$ and the performance of Hit Ratio (HR)@$\{10,20,30\}$. The best results are in bold and the second-best results are underlined. We have the following observations:
\begin{itemize}
    \item On Instacart data, compared with the strongest baseline on each metric, MITGNN improves the Recall, HR, and NDCG on average by $6.85\%$, $0.63\%$, and $6.80\%$, respectively. On Walmart data, MITGNN improves the Recall, HR, and NDCG on average by $16.92\%$, $5.74\%$, and $6.41\%$, respectively. The improvement comes from two perspectives: 1) our model aggregates the neighbors' embeddings to learn the new basket embedding, thus utilizing the item information in the new basket; 2) MITGNN discovers the multi-intent pattern in the basket graph and preserves the semantics in node embeddings, which strengthens the ability to solve BR problems.
    
    \item DREAM performs worst on Walmart data, which suggests that user's shopping patterns are not related to sequential patterns. On Instacart data, DREAM also performs worse than others, except for GC-MC. It implies that sequential patterns are not useful in BR. Since BPR-MF performs much better than DREAM, we should utilize CF signals, rather than sequential patterns. 
    
    \item R-GCN and GC-MC perform worse than BPR-MF. It suggests that direct mean-pooling the neighboring information cannot assist in modeling the useful information for the inductive BR task. NGCF models the interactive information rather than direct mean-aggregation, thus performs better than R-GCN and GC-MC. MITGNN retrieves the interactive information for the BR task, therefore performs the best.
    
    \item Triple2vec directly averages the item embeddings for new baskets, but its performance is worse than BPR-MF, indicating that we need to train a basket aggregator for inductive learning. Our MITGNN trains the embeddings and the aggregators. During inference time of the basket embedding, we aggregate the embeddings of neighbors of the basket. Hence, it performs better.
    
    \item The value of each metric for the inductive BR in Table~\ref{tab:next-basket-Walmart} is higher than the corresponding value for the transductive BR in Figure~\ref{fig:transductive} on Walmart data. The reason is that for inductive BR, we only take out five items for aggregation, and each basket still contains at least $87.5\%$ items as ground truth, compared with only $20\%$ for tranductive BR. Some of these items have already been interacted with users in training baskets, thus are easy to predict. Hence, the values are higher on inductive BR as compared to transductive BR.

\end{itemize}

\subsection{Effect of Intents and Layers}\label{sec:effect_intent}
\begin{figure}
    \begin{subfigure}{.245\textwidth}
    \centering
    \includegraphics[height=0.75\textwidth, width=1\textwidth]{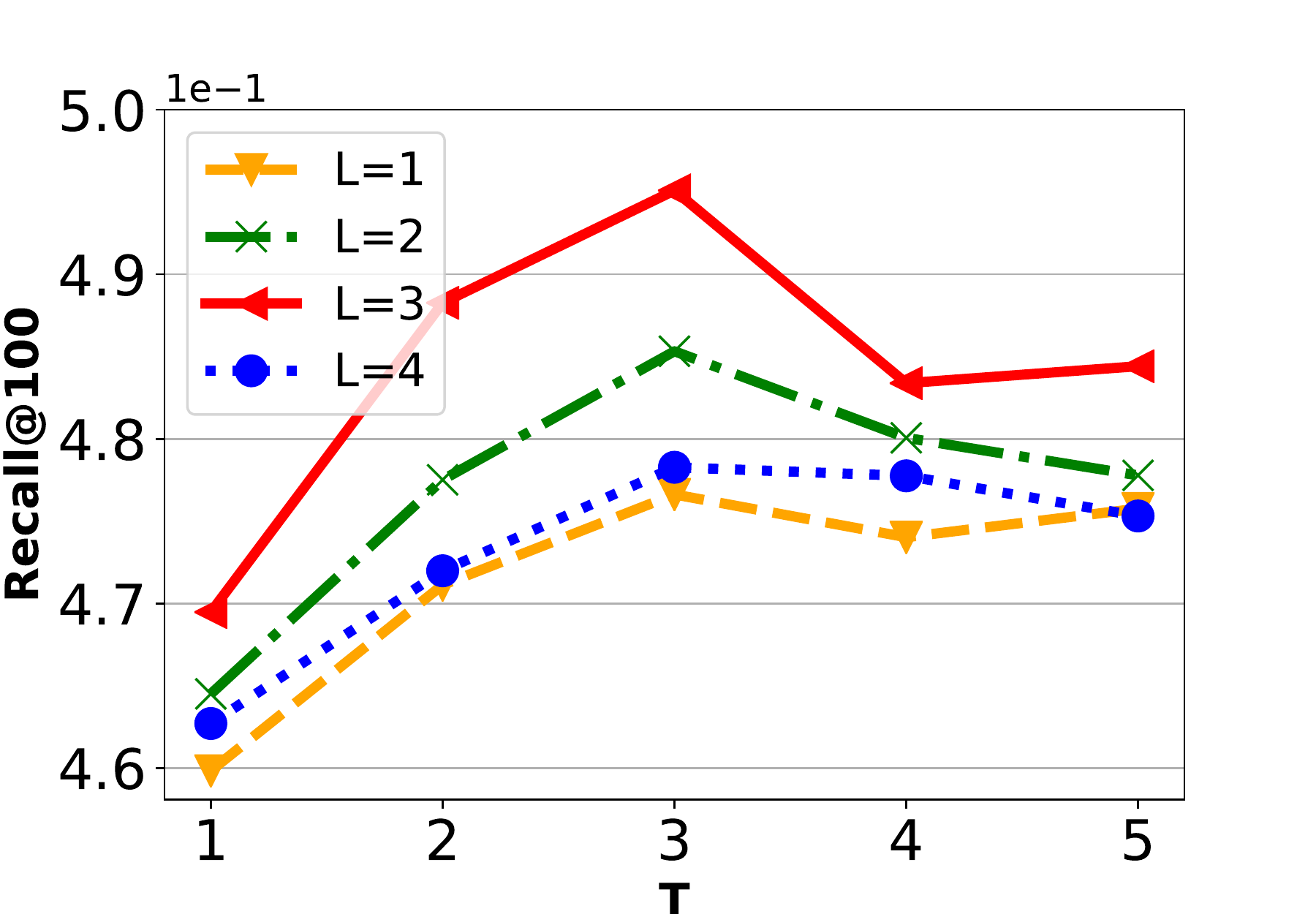}
    \caption{Instacart}
    \label{fig:effect_intent_insta}
    \end{subfigure}
    \hspace{-5mm}
    \begin{subfigure}{.245\textwidth}
    \centering
    \includegraphics[height=0.75\textwidth, width=1\textwidth]{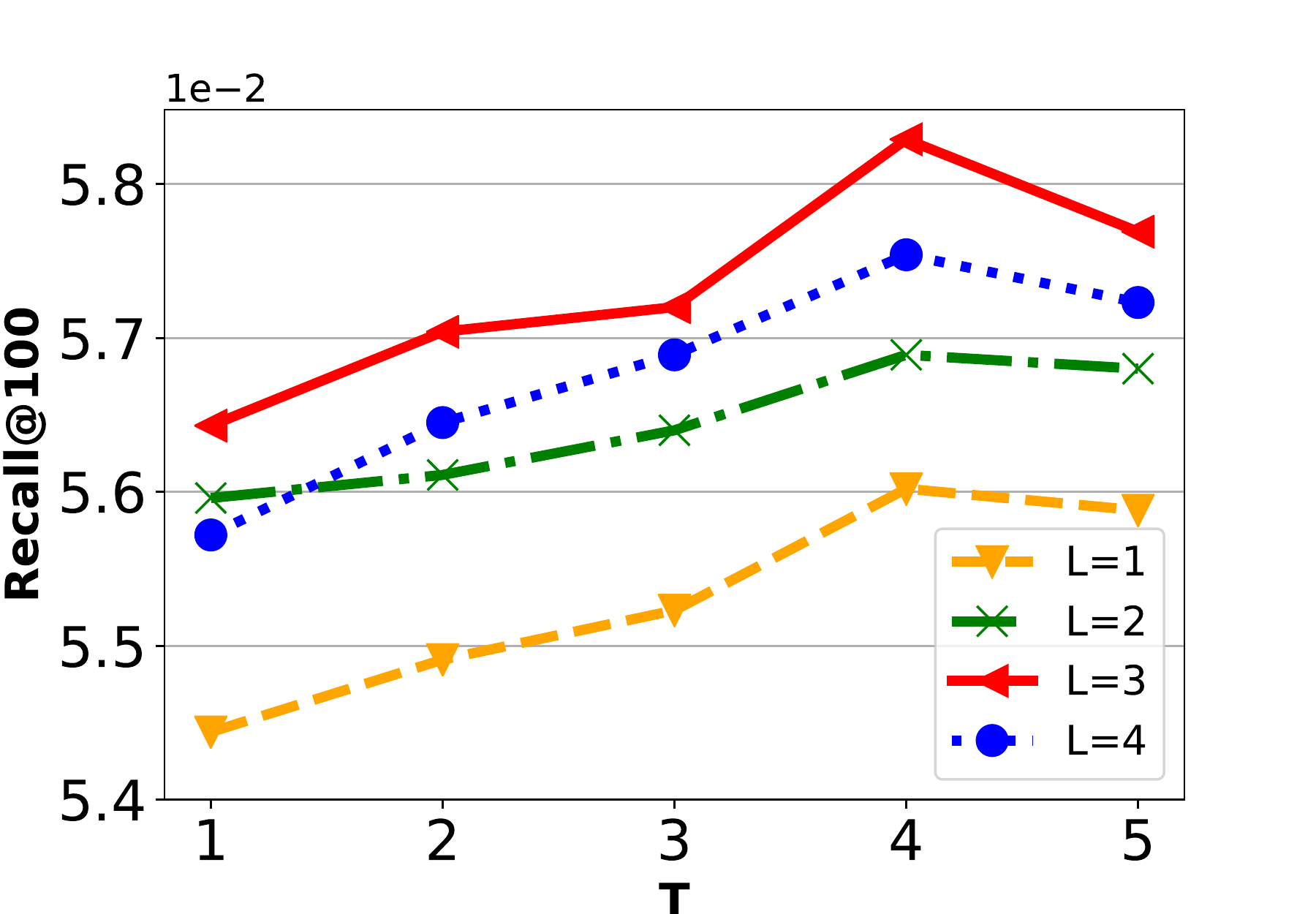}
    \caption{Walmart}
    \label{fig:effect_intent_walmart}
    \end{subfigure}
    \caption{The performance of transductive BR w.r.t. the number of intents $T$ and the number of layers $L$.}
    \label{fig:effect_intent}
\end{figure}

In this section, we discuss the effects w.r.t. the number of the intent $T$ and the number of layers $L$. We discuss the results by presenting the \textit{Recall}@\textbf{100} of transductive BR on two datasets in Figure~\ref{fig:effect_intent}. In each figure, we draw four lines representing the number of layers increasing from $L=1$ to $L=4$. On each line are five points, indicating the number of intents varying from $T=1$ to $T=5$. On both datasets, when the number of intents $T>1$, the performance is always better than $T=1$. This proves that generating multiple hidden intents provides useful information. Moreover, we conclude that we should propagate high-order information in the BR problem, since better evaluation results are obtained when the number of layers $L>1$ as compared to $L=1$.

For the performance on Instacart in Figure~\ref{fig:effect_intent_insta}, we observe that the model consistently outputs the best results when $T=3$, indicating the number of hidden intents on Instacart should be $3$. As we increase the number of intents from $T=1$ to $T=5$, the performance curve goes up and then drops after reaching the peak. We also observe that as the number of layers increases from $L=1$ to $L=3$, performance also increases, which suggests that higher-order signals provide useful information for the BR problem. However, when increasing the number from $L=3$ to $L=4$, the performance drops dramatically, even worse than $L=2$. It is consistent to our choice of the number of layers. 

For the performance on Walmart in Figure~\ref{fig:effect_intent_walmart}, there is gradual improvement when we increase $T$ from $1$ to $4$, which proves the effectiveness of the generated hidden intents. We also observe the results are always the best when the number of intents $T=4$, compared with other numbers of intents. The best number of $T$ for Walmart data is $4$, which is greater than $T=3$ for Instacart. We find that each basket contains more items in Walmart data and is more diverse than Instacart data. It implies more hidden intents are able to model more diversity for baskets. Moreover, $L=3$ generates the highest value against other number of layers. It implies that the higher order provides useful information to the BR problem. Similarly, if we increase the layer number from $3$ to $4$, the performance drops. It implies too much higher-order information spoils the model's ability for BR. 


\section{Conclusion}\label{sec:conclusion}
In this paper, we discuss the multi-intent pattern in BR problems, under both transductive and inductive settings. It is important to model the correlation between intents. Hence, we combine a translation-based model with a GNN model, which is the MITGNN. MITGNN models the intents as vectors translated from basket embeddings. Additionally, MITGNN has a GNN structure with multi-head aggregation to learn the relation vectors between intents and the basket. Moreover, MITGNN has a type-guided attention mechanism to attentively aggregate the hidden intents as user-guided and item-guided basket embeddings for the usage of a user aggregator and item aggregator, respectively. MITGNN is of multi-layer structure and can learn node embeddings by aggregating neighbors. It not only generates the embeddings of entities for transductive BR, but also trains aggregators for inductive BR. Extensive experiments on two large-scale real-world datasets prove the effectiveness of MITGNN on both transductive and inductive BR. Our study on the effect of the number of intents validates the benefits of modeling multi-intent patterns in BR problems. 

\section{Acknowledgements}
This work is supported in part by NSF under grants III-1763325, III-1909323, and SaTC-1930941.

\bibliographystyle{IEEEtran}
\bibliography{references}


\end{document}